\documentclass[aps, jcp, reprint, twocolumn, tightenlines, superscriptaddress, floatfix]{revtex4-1}

\usepackage{graphicx}

\begin{document}

\newcommand{\MACEMP}{\texttt{MACE-MP-0}}
\newcommand{\vk}[1]{{\color{red}{#1}}}

\title{Data-efficient fine-tuning of foundational models for first-principles quality sublimation enthalpies}

\author{Harveen Kaur}
\affiliation{Yusuf Hamied Department of Chemistry,  University of Cambridge,  Lensfield Road,  Cambridge,  CB2 1EW, UK}

\author{Flaviano Della Pia}
\affiliation{Yusuf Hamied Department of Chemistry,  University of Cambridge,  Lensfield Road,  Cambridge,  CB2 1EW, UK}

\author{Ilyes Batatia}
\affiliation{Engineering Laboratory, University of Cambridge, Cambridge, CB2 1PZ UK}

\author{Xavier R. Advincula}
\affiliation{Yusuf Hamied Department of Chemistry,  University of Cambridge,  Lensfield Road,  Cambridge,  CB2 1EW, UK}
\affiliation{Cavendish Laboratory, Department of Physics, University of Cambridge, Cambridge, CB3 0HE, UK}

\author{Benjamin X. Shi}
\affiliation{Yusuf Hamied Department of Chemistry,  University of Cambridge,  Lensfield Road,  Cambridge,  CB2 1EW, UK}

\author{Jinggang Lan}
\affiliation{Department of Chemistry, New York University, New York, NY, 10003, USA}
\affiliation{Simons Center for Computational Physical Chemistry at New York University, New York, New York 10003, USA}

\author{Gábor Csányi}
\affiliation{Engineering Laboratory, University of Cambridge, Cambridge, CB2 1PZ UK}

\author{Angelos Michaelides}
\affiliation{Yusuf Hamied Department of Chemistry,  University of Cambridge,  Lensfield Road,  Cambridge,  CB2 1EW, UK}

\author{Venkat Kapil}
\email{v.kapil@ucl.ac.uk}
\affiliation{Yusuf Hamied Department of Chemistry,  University of Cambridge,  Lensfield Road,  Cambridge,  CB2 1EW, UK}
\affiliation{Department of Physics and Astronomy, University College London, London, UK}
\affiliation{Thomas Young Centre and London Centre for Nanotechnology, London, UK,  London, UK}

\begin{abstract}

Calculating sublimation enthalpies of molecular crystal polymorphs is relevant to a wide range of technological applications.  
However, predicting these quantities at first-principles accuracy -- even with the aid of machine learning potentials -- is a challenge that requires sub-kJ/mol accuracy in the potential energy surface and finite-temperature sampling. 
We present an accurate and data-efficient protocol based on fine-tuning of the foundational \MACEMP{} model and showcase its capabilities on sublimation enthalpies and physical properties of ice polymorphs.
Our approach requires only a few tens of training structures to achieve sub-kJ/mol accuracy in the sublimation enthalpies and sub 1\,\% error in densities for polymorphs at finite temperature and pressure. 
Exploiting this data efficiency, we explore simulations of hexagonal ice at the random phase approximation level of theory at experimental temperatures and pressures, calculating its physical properties, like pair correlation function and density, with good agreement with experiments. 
Our approach provides a way forward for predicting the stability of molecular crystals at finite thermodynamic conditions with the accuracy of correlated electronic structure theory.

\end{abstract}

\maketitle

\section{Introduction}
\label{s:intro}

Molecular crystals form an essential class of materials with technological applications in industries such as pharmaceuticals~\cite{bernstein_pinching_2005}, electronics~\cite{forrest_path_2004}, and agriculture~\cite{yang_ddt_2017}. 
Often, molecular crystals exhibit competing polymorphs, i.e., multiple metastable crystalline phases with very similar stability (for instance, relative free energies can be within $\approx$ 1\,kJ/mol error)~\cite{price_predicting_2014}. 
While the most common experimental probe of the polymorph stability is the sublimation enthalpy, recent work shows discrepancies across calorimetry literature for prototypical molecular crystals beyond 1\,kcal/mol~\cite{della_pia_how_2024} $\approx$ 4.2 \,kJ/mol.
Hence, there is a need for an independent estimation of sublimation enthalpies using first-principles methods. \\

Although possible in theory, predicting sublimation enthalpies with first principles methods is challenging due to the need for high accuracy.
Reliable predictions require a tolerance of nearly 1\,kcal/mol for absolute sublimation enthalpies and a tighter tolerance of less than 1\,kJ/mol for relative sublimation enthalpies~\cite{price_predicting_2014}.
As shown by ~\citet{zen_fast_2018}, predicting absolute sublimation enthalpies of common molecular solids, such as ice, ammonia, carbon dioxide and aromatic hydrocarbons, consistently to 1\,kcal/mol requires computationally demanding ``correlated" electronic structure techniques.
These techniques include quantum fixed-node Diffusion Monte Carlo~\cite{foulkes_quantum_2001}, periodic coupled cluster~\cite{booth_towards_2013} or random phase approximation (RPA) with singles excitations~\cite{klimes_lattice_2016, ren_beyond_2011}.
Similarly, achieving correct relative stabilities, accurate 1\,kJ/mol, of prototypical polymorphs of oxalic acid, glycine, paracetamol, and benzene requires statistical mechanics incorporating dynamical disorder via thermal effects~\cite{rossi_anharmonic_2016, kapil_assessment_2019}, thermal expansion~\cite{kapil_complete_2022}, and anharmonic quantum nuclear motion~\cite{rossi_anharmonic_2016, kapil_assessment_2019, kapil_complete_2022}. \\

Unfortunately, the computational cost associated with correlated electronic structure theory~\cite{zen_fast_2018} or rigorous quantum statistical mechanics~\cite{markland_nuclear_2018}, individually or in tandem, remains high.
Hence, sublimation enthalpies are commonly approximated inexpensively with dispersion-corrected density functional theory (DFT) for a static geometry optimized lattice at zero kelvin~\cite{hoja_reliable_2019}.  
As a consequence of their static description, these enthalpies are compared indirectly with experiments requiring a careful extrapolation of measured enthalpies to a static lattice at zero kelvin~\cite{whalley_energies_1984}.
Unfortunately, this is typically associated with an error-prone \textit{ad hoc} subtraction of zero-point energy corrections calculated at DFT~\cite{dolgonos_revised_2019}.
Furthermore, considering the inherent uncertainties in measured sublimation enthalpies~\cite{della_pia_how_2024}, there is a need for relative stabilities that can be unambiguously compared with experiments at their respective thermodynamic conditions~\cite{kapil_complete_2022}. \\

In this context, machine learning potentials (MLPs)~\cite{behler_generalized_2007, bartok_gaussian_2010, drautz_atomic_2019, batzner_e3-equivariant_2022, batatia_mace_2022} provide an avenue for first-principles-accuracy modelling of molecular crystals at finite temperature.
MLPs have been used as computationally inexpensive surrogates for first-principles potential energy surfaces (PES) for ranking putative polymorphs in increasing order of lattice energies~\cite{deringer_data-driven_2018, wengert_data-efficient_2021,butler_machine_2023}. 
MLPs have also facilitated finite-temperature modelling of polymorphs of simple compounds like hydrogen~\cite{cheng_evidence_2020} and water~\cite{cheng_phase_2021, zhang_phase_2021}, with converged system sizes and simulation times.
More recently, ~\citet{kapil_complete_2022} developed an MLP-based framework for predicting polymorph relative stabilities for paradigmatic molecular crystals containing up to four chemical species, such as benzene, glycine, and succinic acid. 
While their approach enables rigorous predictions of relative and absolute stabilities at finite temperatures, it presents a number of limitations arising from the limitations of conventional MLPs. 
These include a $>$ kJ/mol error in out-of-distribution prediction, the need for large volumes of training data ( $> $1000 structures per compound), and declining accuracy and data efficiency with an increasing number of chemical species.
These deficiencies limit predictive finite-temperature stability calculations for generic molecular compounds using chemically accurate electronic structure theory level~\cite{zen_fast_2018}. \\

In this work, we present a highly accurate and data-efficient MLP-based approach for finite-temperature modelling and sublimation enthalpy prediction of given polymorphs of a compound.
Using ice polymorphs as a test bed, we show that using the multi atomic cluster expansion (MACE) architecture~\cite{batatia_mace_2022} supplemented with fine-tuned training of the foundational \MACEMP{} model, is sufficient to reach sub-kJ/mol accuracy with as few as 50 training structures.
In Section~\ref{s:methods}, we discuss the shortcomings of conventional MLPs and the capabilities of the newer methods with a focus on MACE and \MACEMP{}, followed by the details of our protocol, including the dataset generation, training and validation steps.
In Section~\ref{s:results}, we apply our approach to crystalline ice -- a prototypical system exhibiting a high degree of polymorphism with good quality experimental data on densities and sublimation enthalpies~\cite{whalley_energies_1984}.
We demonstrate the accuracy and generality of our approach on the excellent agreement of finite-temperature density and sublimation enthalpies of ice polymorphs directly against DFT level. 
Finally, with a few tens of periodic RPA calculations, we simulate the physical properties of ice Ih at finite temperature and pressure in good agreement with experimental data.
In Section~\ref{s:conclusions}, we discuss future efforts for direct finite-temperature simulations for characterizing molecular crystal polymorphs at the accuracy of correlated electronic structure.

\section{Theory and methods}
\label{s:methods}

\subsection{Brief review of machine learning models \label{ss:MLPs}}

MLPs typically represent the total energy of a system as a sum of atomic energies. 
Standard models of the atomic energy of a central atom first preprocess the relative atomic positions of all atoms up to a cutoff into so-called ``atomic representations"~\cite{musil_physics-inspired_2021, deringer_gaussian_2021}.
Subsequently, the representations are used as inputs to a regression model~\cite{musil_physics-inspired_2021, deringer_gaussian_2021}, such as a Gaussian process~\cite{bartok_gaussian_2010}, and artificial~\cite{behler_generalized_2007} or deep neural networks~\cite{zhang_deep_2018}, trained on total energy of the system and its gradients such as atomic forces and virial tensors~\cite{bartok_representing_2013, behler_constructing_2015}. 
Typically, these representations are $n$-body correlation functions (defined for every $n$-tuple of atom types and typically truncated at $n$=2 or 3) of relative atomic positions which encode rotational, permutational, and inversion invariances~\cite{willatt_atom-density_2019, drautz_atomic_2019}. \\

Standard architectures, such as the Behler-Parrinello neural network (BPNN) framework~\cite{behler_constructing_2015}, Gaussian Approximation Potential~\cite{bartok_gaussian_2010}, SchNet~\cite{schutt_schnet_2018}, DeepMD~\cite{zhang_deep_2018}, and Moment Tensor Potential~\cite{shapeev_moment_2016}, can be constructed by mixing and matching various flavours of two- or three-body atomic representations and regression models.
Despite their success, standard models have two main limitations. 
First, the truncation of body order leads to the incompleteness of the atomic representations, limiting their accuracy and smoothness~\cite{pozdnyakov_incompleteness_2020}.
The accuracy can be systematically improved, but including higher body order representations involves a much higher computational cost and labour~\cite{nigam_recursive_2020, bigi_wigner_2023, shapeev_moment_2016}.
Second, the number of representations scales combinatorially with the number of chemical species as $n$-body representations are defined for every $n$-tuple of atomic species. 
Hence, for a given accuracy cutoff, standard MLPs exhibit an exponentially increasing cost with an increasing number of chemical species.
Similarly, for a fixed cost, these models have a steeply worsening accuracy with increasing chemical species. \\
Newer MLPs such as NequIP~\cite{batzner_e3-equivariant_2022}, MACE~\cite{batatia_mace_2022}, and PET~\cite{pozdnyakov_smooth_2024}, implemented as Euclidean graph neural networks ({\em e3nn})~\cite{geiger_e3nn_2022}, address these issues by incorporating (near) complete atomic representations~\cite{drautz_atomic_2019}.
In addition, they exploit a learnable latent chemical space~\cite{willatt_feature_2018, batzner_e3-equivariant_2022, batatia_design_2022} for smoothly interpolating or extrapolating representations across chemical species at $\mathcal{O}(1)$ cost without compromising accuracy~\cite{lopanitsyna_modeling_2023}.
Specifically, in the MACE architecture, the initial node representations of the graph neural network are based on the atomic cluster expansion~\cite{drautz_atomic_2019} up to a selected body order (typically $n=4$).
MACE systematically constructs higher body-order representations in terms of the output of the previous layer and an equivariant and high body-order message passing scheme~\cite{batatia_design_2022}. 
Hence, increasing the number of layers or body order of the message passing enables learning correlation functions of arbitrary order. 
However, thanks to its message passing scheme, MACE can efficiently construct high body-order representations with a simple architecture (e.g., the default parameters of MACE with just two layers give access to a body order of 13)~\cite{kovacs_evaluation_2023}.
Finally, embedding chemical information in a learnable latent space~\cite{darby_tensor-reduced_2023}, MACE displays an $\mathcal{O}(1)$ computational cost with the number of chemical species. \\

Exploiting these capabilities for datasets with a large number of elements, ~\citet{batatia_foundation_2024} have recently developed a so-called \MACEMP{} foundational MLP model trained on a diverse Materials Project dataset.
Specifically, \MACEMP{} is trained on the \texttt{MPTrj} dataset, comprising 1.5 million small periodic unit cells of inorganic (molecular) crystals with elements across the periodic table. 
The training set includes total energies, forces and stress tensors estimated at the PBE(+U) level.
Trained for elements across the periodic table, the \MACEMP{} model is capable of out-of-the-box usage for general materials with qualitative (and sometimes quantitative) PBE accuracy.
Other classes of foundational MLPs exist, such as \texttt{CHGNET}~\cite{choudhary_unified_2023} and \texttt{M3GNet}~\cite{chen_universal_2022}, based on materials project datasets and the \texttt{ALIGNN-FF}~\cite{choudhary_unified_2023} based on the \texttt{JARVIS-DFT} dataset. 
Unlike \MACEMP{}, these models are based on three-body atomic representations. \\

While \MACEMP{}'s accuracy is insufficient for studying molecular crystal polymorphs, its parameters may provide a starting point for training system-specific models at a different level of theory.
Considering that the pre-trained $n$-body atomic representations are valid for generic materials, using the \MACEMP{} parameters as a starting point for fine-tuning may require less data and computational time compared to training a new model from scratch.

\subsection{Details of our framework \label{ss:dataset}}

\subsubsection{The protocol}

We propose a simple pipeline for studying the physical properties of a given polymorph using MLPs at a desired thermodynamic state point (temperature $T$ and pressure $P$). It includes the following steps.

\begin{enumerate}
    \item \textbf{Dataset Sampling:} We perform a short first-principles molecular dynamics (MD) simulation in the $NPT$ ensemble. To ensure this step is inexpensive, we select a generalized gradient approximation (GGA) DFT level of theory and coarsely converged electronic structure parameters and simulation lengths up to 5\,ps. 

    \item \textbf{Dataset Generation}: We randomly select 500 structures to perform total energy, force, and stress calculations with tightly converged parameters. We collect total energies, forces and stress tensors as target properties, thereby generating a dataset of 500 structures, energy and its gradients. The validation set includes 100 randomly selected structures. The remaining structures are split into training sets of increasing sizes with 50, 100, 160, 200, 320 and 400 structures. The larger sets include structures from the smaller ones. 
    
    \item \textbf{Model development and validation}: We train two types of models for each training set --  a MACE model trained from scratch and a MACE model fine-tuned from \MACEMP{} in which we use the initial parameters from the foundational models as a starting point. We compare learning curves by plotting the root mean square errors (RMSEs) of the total energies and the atomic force components, thereby identifying the training set sizes and the training methods that deliver sub-kJ/mol accuracy.
    
    \item \textbf{Model testing against DFT}: We perform an additional out-of-distribution test in the $NPT$ ensemble sampled by the MLPs and compare it against the DFT ensemble. We study the convergence of the average potential energy and density from $NPT$ simulations as a function of the size of the training set. We obtain a converged DFT reference for the average potential energy and density by performing DFT single point energy calculations on $100$ uniformly stridden configurations. We use the difference between MACE and DFT energies to estimate the averages at DFT level using statistical reweighting. We identify the training set sizes that deliver good accuracy against the DFT reference ensemble.
        
    \item \textbf{Upgrade (optional)}: The smallest training set identified in step 4 may be upgraded to a higher level of theory, such as a hybrid-functional DFT level or an explicitly correlated electronic structure theory level with gradient implementation, such as the random phase approximation (RPA).

\end{enumerate}

The models from step 4 (or 5) can be used for production simulations in the $NPT$ ensemble with larger simulation sizes and long simulation times.

\subsubsection{Computational details}

We organize the technical details as follows: \\

\textbf{Systems and Thermodynamic conditions}: We validate our pipeline on the densities and sublimation enthalpies of ice polymorphs Ih, II, VI and VIII at 100\,K and 1 bar. 
We use simulation cells with 128 molecules for ice Ih, 96 molecules for ice II and 80 molecules for ice VI and VIII.
These simulation cells ensure lattice parameters greater than $10\,\AA{}$. 
We employ the revPBE functional with (zero-damping) D3 dispersion correct due to its good performance against diffusion Monte Carlo for ice phases~\cite{della_pia_dmc-ice13_2022}.  \\

\textbf{Dataset Sampling}: We use the \texttt{CP2K} code~\cite{kuhne_cp2k_2020} for efficient sampling of the dataset via ab initio molecular dynamics simulations. The electronic structure is described using Kohn-Sham density functional theory with a plane-wave basis set truncated at an energy cutoff of 500 Rydberg, TZV2P-GTH basis sets~\cite{vandevondele2007gaussian}, GTH-PBE pseudopotentials~\cite{goedecker1996separable}, and $\Gamma$-point sampling of the Brillouin zone. The simulations are carried out in the NPT ensemble using an isotropic cell at a constant pressure of 1 bar and 100\,K. \\

\textbf{Dataset Generation}: We use the \texttt{VASP} code~\cite{VASP2, VASP3, VASP4} to perform single-point energy calculations at revPBE-D3 level with SCF parameters from Ref.~\citenum{della_pia_dmc-ice13_2022}. Our over-conservative parameters allow us to minimize the amount of noise in our training set. We used hard PAW (PBE) pseudopotentials~\cite{VASP_PAW1, VASP_PAW2} with an energy cutoff of $1000 \text{ eV}$, $\Gamma$-point with supercells lattice parameters exceeding 10$\AA{}$, and a dense FFT grid (PREC=High). \\

\textbf{Model training (from scratch)}: We use two MACE layers, with a spherical expansion of up to $l_{\text{max}}=3$, and 4-body messages in each layer (correlation order 3).
We use a self-connection for both layers, a 128-channel dimension for tensor decomposition and a radial cutoff of 6\AA\. 
We expand the interatomic distances into 8 Bessel functions multiplied by a smooth polynomial cutoff function to construct radial features, which in turn fed into a fully connected feed-forward neural network with three hidden layers of 64 hidden units and SiLU non-linearities.
A maximal message equivariance of $L=2$ is applied. 
The irreducible representations of the messages have alternating parity (in {\em e3nn} notation, $\texttt{128x0e + 128x1o + 128x2e}$).\\

\textbf{Model training (fine-tuned from \MACEMP{})}: 
We fine-tune the large \MACEMP{}~\cite{batatia_foundation_2024} model by continuing training from the last checkpoint and, therefore, using the same hyperparameters.
A self-connection is used only at the first layer. The remaining parameters are the same as the model trained from scratch. \\ 

\textbf{Model testing}: We use the \texttt{i-PI} code~\cite{kapil_i-pi_2019} to perform $NPT$ MD simulations using an \texttt{ASE}~\cite{larsen_atomic_2017} as a client for calculating MACE total energy, forces and virials. We perform 50\,ps long simulations employing a timestep of 0.5\,fs. We use the fully flexible Martyna-Tuckerman-Tobian-Klien~\cite{martyna_explicit_1996} barostat implementation with a relaxation time of 1\,ps. For efficient sampling, we use an optimally damped generalized Langevin equation thermostat for the system and the lattice degrees of freedom~\cite{ceriotti_langevin_2009}. We sample positions, potential energies, and densities every 100 MD steps. \\

\textbf{Upgrade}: For ice Ih, we upgrade the electronic structure theory to RPA@PBE0, wherein the RPA correlation energy is computed using the hybrid functional PBE0 as the starting point. We employ the sparse tensor-based nuclear gradients of RPA ~\cite{bussy2023sparse} as implemented in \texttt{CP2K}. All calculations use the triple-zeta cc-TZ and RI\_TZ basis sets alongside GTH-PBE0 pseudopotentials. The Hartree-Fock exchange contribution to the SCF and the Z-vector equation is calculated within the ADMM approximation~\cite{guidon2010auxiliary},
utilizing the admm-tzp auxiliary basis.\\

\section{Results and Discussion}
\label{s:results}

\subsection{Performance on validation set}

\begin{figure}[!h]
    \centering
    \includegraphics[width=0.5\textwidth]{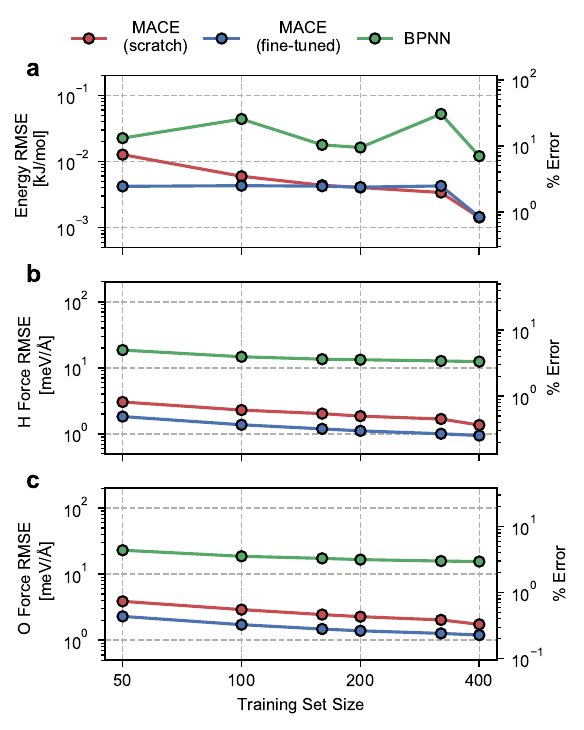}
    \caption{\textbf{Root mean square errors for ice Ih on the validation set as a function of the number of training structures.} Panel \textbf{a}, \textbf{b}, and \textbf{c} respectively report the error in the energy per water molecule, the mean force on hydrogen atoms, and the mean force on oxygen atoms. We also report relative RMSEs that correspond to the \% error relative to the standard deviation of the quantities in the validation set. Red, blue and green circular markers correspond to errors for MACE models trained from scratch, MACE models fine-tuned from the \MACEMP{} model, and a Behler-Parrinello Neural Network Potential, respectively. Coloured lines are a guide for the eye.}
    \label{fig:learning_curves}
\end{figure}

We begin by evaluating the accuracy and data efficiency of the MACE architecture and the ``from scratch" and ``fine-tuned" training protocols. 
We perform this test on the ice Ih phase and check whether the MACE models deliver sub-kJ/mol accuracy on the validation set. \\
As shown in Fig.~\ref{fig:learning_curves}(a), we report RMSEs of the energy per molecule and the force components on H and O atoms as a function of the size of the training set. 
A MACE model trained from scratch reports an energy RMSE of nearly 0.01\,kJ/mol with the smallest training set comprising 50 structures. 
On an absolute scale, this error is extremely low. 
However, it corresponds to 10\% of the mean energy variation in the validation set, which is nearly the per atom standard deviation of the potential energy at 100\,K.
With 400 structures, i.e., the entire dataset, we report an energy RMSE of around 0.001 kJ/mol. 
This error corresponds to around 1\% of the validation set standard deviation.  \\

We next study the effect of fine-tuning the parameters of the \MACEMP{} model, which implies starting the training from the last checkpoint of the foundational model.
As seen in Fig.~\ref{fig:learning_curves}(a), fine-tuning improves the accuracy of the models, compared to training from scratch, for small datasets containing fewer than 160 structures. 
With just 50 training structures, a fine-tuned MACE model reports an energy RMSE that corresponds to around 2\% of the validation set standard deviation. 
For training set sizes, beyond 160 structures we see nearly identical performance of the from scratch and fine-tuned models on the energy RMSE. \\

The RMSEs of the force components on H and O atoms paint a similar picture.
The from-scratch MACE models report a small RMSE of nearly 2 mev$/\AA{}$ with just 50 structures, nearly 1\% of the validation set standard deviation, with a systematic reduction in error with increasing training data.
On the other hand, fine-tuning the \MACEMP{} model results improved accuracy and data efficiency with nearly 1\,meV$/\AA{}$ RMSE with 50 structures and sub meV$/\AA{}$ RMSE with over 200 structures. 
We observe improvements in force RMSEs across the full range of training set sizes.  \\

To contextualize these RMSEs, we also report RMSEs obtained with a standard 2- or 3-body atomic-representation-based model. 
We select the BPNN~\cite{behler_constructing_2015} architecture which has been widely used for simulating the bulk~\cite{cheng_ab_2019}, interfacial~\cite{kapil_first-principles_2023} and confined phases of water~\cite{kapil_first-principles_2022}. 
We observe nearly order-of-magnitude higher energy and force RMSEs with the BPNN scheme, with indications of saturating errors with 400 training structures.
Although on an absolute scale, the BPNN model reports small energy RMSE, it corresponds to saturation at around 10\% relative error.
The saturation in the RMSEs is likely due to a ceiling on learning capacity due to incomplete atomic representations. 
We note the much higher data efficiency of the MACE models, which exhibit a higher accuracy even with an order of magnitude and fewer training data.

\subsection{Performance at finite temperature and pressure}

\begin{figure}
    \centering
    \includegraphics[]{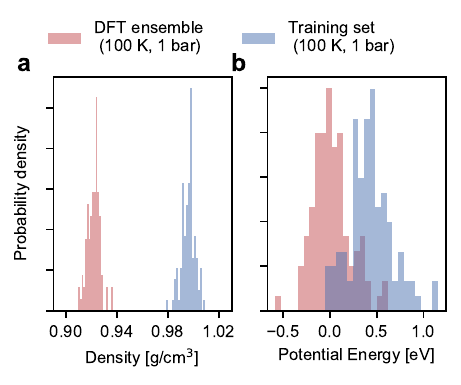}
    \caption{\textbf{Distribution of density and volume in the training and target ensembles.} Panels \textbf{a} and \textbf{b} respectively show the histograms of the density and the potential energy of ice Ih in the DFT and the training set ensembles. 
    The training set histograms are estimated for 100 configurations randomly sampled from a 5\,ps long first-principles MD simulation with unconverged DFT parameters. The potential energies are subsequently reevaluated with converged DFT parameters. The DFT ensemble histograms are estimated via statistical reweighting using configurations from a 5\,ps fine-tuned MACE simulation trained on 100 structures (see main text). For clarity, we realign the energies to the median of the DFT ensemble energies.}      \label{fig:ensembles}
\end{figure}

As shown in Fig.~\ref{fig:ensembles}, the configurational ensemble used to generate the training and validation datasets deviates significantly in energy and volume distributions from that of revPBE-D3 with converged electronic structure parameters.
The configurations used for training correspond to denser structures, with shorter interatomic distances and, consequently, higher potential energies compared to the DFT ensemble.
Hence, the RMSEs in Fig~\ref{fig:learning_curves} only reflect the quality of regression in the training ensemble.  \\

\begin{figure}
    \centering
    \includegraphics[]{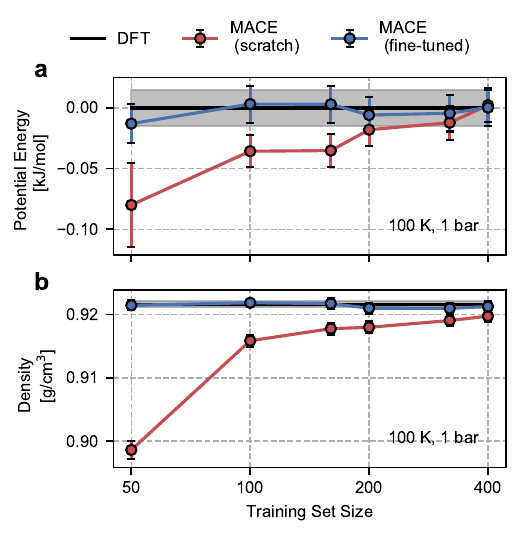}
    \caption{\textbf{Finite-temperature testing for ice Ih as a function of the volume of training data.} Panels \textbf{a} and \textbf{b} respectively show the thermodynamic average of the potential energy and density in the $NPT$ ensemble at 1\,bar and 100\,K. Circular markers in red and blue respectively correspond to data generated by MACE models trained from scratch and MACE models fined-tuned from the \MACEMP{} model. Coloured lines are a guide to the eye. The black lines correspond to DFT thermodynamic averages estimated by statistical reweighting using configurations from a fine-tuned MACE model. The grey region corresponds to a $1\,\sigma$ error statistical error estimated from block averaging. \label{fig:finite_temperature_validation}}
\end{figure}

To assess model performance at finite thermodynamic conditions, we perform fully flexible $NPT$ simulations at 100\,K and 1\,bar for each model. 
We found the BPNN $NPT$ simulations to be unstable due to overfitting on the training ensemble, hence we only present results for the MACE models. 
We report the thermodynamic average of the potential energy and the density as a function of the size of the training set in Fig.~\ref{fig:finite_temperature_validation}.
Exploiting MACE's high fidelity, we perform statistical reweighting to calculate the DFT reference of the average potential energy and the density.
For this purpose, we use the trajectory sampled by the fine-tuned MACE trained on 100 structures.
The DFT references allow us to test MLPs against their DFT in their thermodynamic ensembles directly. \\

\begin{table*}[!ht]
    \centering
    \begin{tabular}{ c || c | c |  c || c |  c | c}
    \hline
    Polymorph & \multicolumn{3}{c||}{density [g/cm$^{3}$]}  & \multicolumn{3}{c}{ sublimation enthalpy [kJ/mol]} \\
    \hline
        &   MACE & DFT & \% Error &    MACE & DFT & Error\\
    \hline
    Ih  &   0.922 $\pm$ 0.001 & 0.922 $\pm$ 0.001 & 0.012 $\pm$ 0.161 & -58.208 $\pm$ 0.089 & -58.214 $\pm$ 0.089 & 0.006 $\pm$ 0.126\\
    II  &   1.200 $\pm$ 0.002 & 1.190 $\pm$ 0.009 & 0.896 $\pm$ 0.764 & -57.055 $\pm$ 0.091 & -56.814 $\pm$ 0.145 & 0.241 $\pm$ 0.0172\\
    VI  &   1.329 $\pm$ 0.001 & 1.330 $\pm$ 0.006 & 0.083 $\pm$ 0.461 & -55.225 $\pm$ 0.091 & -55.215 $\pm$ 0.114 & 0.010 $\pm$ 0.146 \\
    VIII    &   1.569 $\pm$ 0.001 & 1.565 $\pm$ 0.003 & 0.262 $\pm$ 0.196 & -56.063 $\pm$ 0.107 & -55.719 $\pm$ 0.143 & 0.349 $\pm$ 0.179\\
    \hline
    \end{tabular}
    \caption{\textbf{Physical properties of ice polymorphs at 100\,K and 1\,bar}. Densities and sublimation enthalpies of ice Ih, II, VI, and VIII are estimated in the $NPT$ ensemble for the MACE potentials and the underlying DFT level. The discrepancy of MACE with respect to DFT is expressed as a percentage error. Uncertainties correspond to 1\,$\sigma$ standard errors of the mean. \label{tab:ice_polymorphs_physical_properties}}
\end{table*}

As shown in Fig.~\ref{fig:finite_temperature_validation}, the from-scratch MACE models generalize well to the true thermodynamic ensemble displaying sub-kJ/mol error for the average potential energy at 100\,K and 1\,bar.
Despite these small errors, we note that the from-scratch MACE thermodynamic averages deviate from the DFT reference at 50 structures, yielding a statistically significant agreement for models trained with more than 200 structures.
These small errors lead to significant disagreements in the density (a quantity that is much harder to converge compared to energy or forces as per empirical evidence~\cite{magdau_machine_2023}). 
We require training on 400 structures to converge the density to the DFT reference within statistical error. \\

With the fine-tuned models, we observe a remarkable performance against the DFT ensemble. 
Even for the smallest training set comprising 50 structures, we observe a quantitative agreement in the density and the average energy. 
This is likely a consequence of the pre-trained atomic representations of \MACEMP{}~\cite{batatia_foundation_2024} that describe the general volume dependence of the energy. 
These tests suggest the potential of our approach for finite-temperature modelling of molecular polymorphs with fewer than 100 training structures.
This is a marked improvement over our previous work in Ref.~\cite{kapil_complete_2022}, which required over a few hundred or thousands of structures and differential learning for stable $NPT$ simulations. 

\subsection{DFT-level sublimation enthalpies and physical properties of ice polymorphs}

The fine-tuned MACE model only requires up to a hundred training structures for ice Ih for first-principles-quality $NPT$ simulations.
We next check if the observed data-efficiency for ice Ih is valid for other polymorphs. 
For this purpose, we predict the sublimation enthalpy and density of ice II, VI, and VIII at finite thermodynamic conditions and check agreement with the DFT ensemble. \\

For each polymorph, we train a fine-tuned MACE model on 100 structures and perform $NPT$ simulations at 100\,K and 1\,bar to estimate the density and the average potential energy.
To estimate the sublimation enthalpy, we further train a fine-tuned MACE model on 200 structures of a water molecule providing a gas phase reference enthalpy at 100\,K and 1\,bar. 
The training set was developed in the same way as for the ice polymorphs with initial $NVT$ sampling using \texttt{CP2K} at revPBE-D3 level and converged DFT calculations on randomly sampled structures using \texttt{VASP}. 
Finally, for an apples-to-apples comparison, we compare with DFT-level densities and sublimation enthalpies calculated using statistical reweighting. \\

As shown in Table~\ref{tab:ice_polymorphs_physical_properties}, the densities and sublimation enthalpies estimated with MACE at 100\,K and 1\,bar agree remarkably with the reference DFT estimations up to the statistical error. 
In most cases, the discrepancies between MACE and DFT are within the $1\,\sigma$ statistical error of DFT estimations. 
In all the cases, the agreement for the sublimation enthalpy and the density are within $0.5\,$kJ/mol and $1\,\%$ respectively.
These results demonstrate the data efficiency and generalizability of the fine-tuned MACE models to the full $NPT$ ensemble of energy and volume despite being trained on a skewed ensemble. 

\subsection{Random phase approximation level physical properties of ice Ih in the $NPT$ ensemble}

The sub-kJ/mol accuracy of the fine-tuned MACE models at finite thermodynamic conditions suggests that they are capable of learning at the accuracy of correlated electronic structure theory levels for relative and absolute sublimation enthalpies. 
In addition, the low data requirement of the fine-tuned MACE models makes them practically viable for training on small datasets generated at computationally demanding correlated electronic structure theory level.
To demonstrate this, we trained an RPA-level fine-tuned MACE model for ice Ih using 75 periodic RPA total energy and force single point calculations estimated with \texttt{CP2K}. \\

We found that training at RPA level was more challenging than at DFT level. 
Our training set resulted in an energy mean absolute error (MAE) of 0.03 kJ/mol and a force MAE of 24.8 mev/$\AA{}$, which is significantly higher than the DFT-level energy MACE of 0.001 meV/atom and force 1 mev/$\AA{}$ MAE on the training set. 
This is due to the noise in the RPA total energies and forces resulting from the resolution of identity triple zeta basis sets~\cite{bussy2023sparse} and the auxiliary density matrix method~\cite{guidon2010auxiliary}.
We confirmed the noise in RPA forces by noting that the sum of the forces on all atoms is non-zero and averages to around 60 meV/$\AA{}$. \\

\begin{figure*}
    \centering
    \includegraphics[]{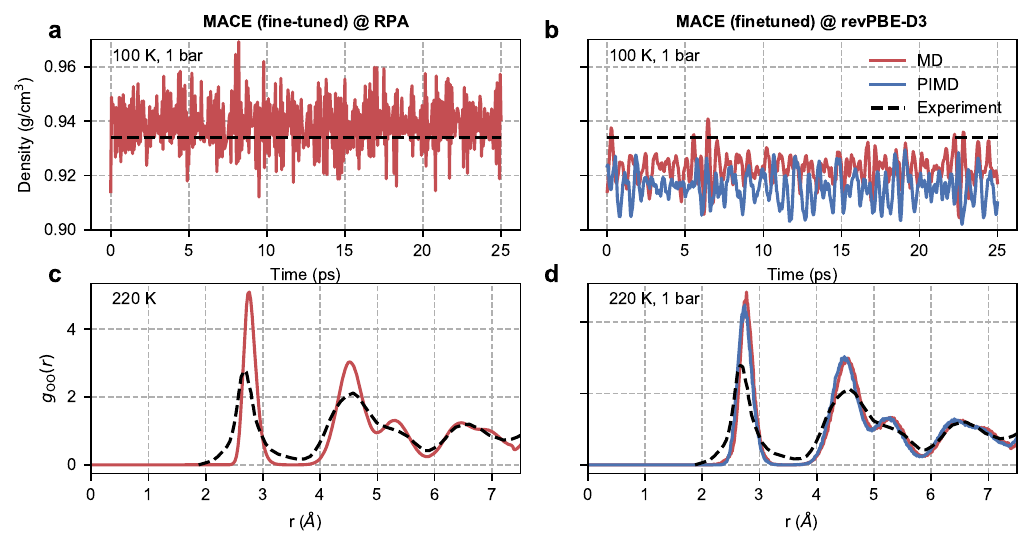}
    \caption{\textbf{RPA-level simulations of ice Ih}. Panel \textbf{a} shows the time series of the density of ice Ih from a classical $NPT$ simulation at 100\,K and 1\,bar with a MACE model trained at RPA level.  Panel \textbf{b} shows the same quantity but with classical and path-integral $NPT$ simulations using the MACE model trained at the revPBE-D3 level. The black dashed lines correspond to the experimental density~\cite{lide2005crc} at 100\,K and 1\,bar. Panel \textbf{c} shows the oxygen-oxygen pair correlation function of ice Ih at 220\,K estimated from an RPA level MACE $NVT$ simulation. Panel  \textbf{d} reports the same quantity but with classical and path-integral $NPT$ simulations at 220\,K and 1\,bar using the MACE model trained at revPBE-D3 level. The black lines correspond to the experimental pair correlation function~\cite{soper_water_2014} measured at 220\,K and 1\,bar.}
    \label{fig:rpa_simulations_ice_Ih}
\end{figure*}

Despite the noisy training set, we were able to report stable $NPT$ simulations at RPA level at 100\,K and 1\,bar, as shown in Fig.~\ref{fig:rpa_simulations_ice_Ih}. 
However, due to the high cost of RPA single-point calculations, we were unable to perform statistical reweighting for direct validation. 
Nonetheless, the accuracy afforded by RPA simulation allowed us to compare meaningfully with experiments. \\

Our $NPT$ simulations reported a density of 0.939 g/cm$^3$, which is slightly higher than the experimental density of 0.934 g/cm$^3$~\cite{lide2005crc}, but a marked improvement over revPBE-D3 density of 0.922 g/cm$^3$. 
The missing quantum nuclear effects in our RPA simulations explain the remaining (small) discrepancy between the RPA $NPT$ and experimental densities. 
We were unable to confirm this directly as our path integral simulations were not stable.
However, we were able to confirm that the instability of the simulations is linked to the noisy fit to RPA forces by performing path integral simulations using our revPBE-D3 level fine-tuned model. 
As can be seen in Fig.~\ref{fig:rpa_simulations_ice_Ih}(b), we report stable simulations with quantum nuclear effects despite not training on configurations generated using path integral simulations.
Quantum nuclear effects marginally reduce the density of ice Ih, nearly to the same extent as the overestimation of the classical RPA density compared to experiments. \\ 

Finally, with access to stable trajectories, we compared the structure of ice Ih with radiation total scattering experiments at 220\,K~\cite{soper_water_2014} by calculating the oxygen-oxygen pair correlation function.
Although our potential is trained on configurations corresponding to a 100\,K and 1\,bar ensemble, our models can generalize to higher temperatures.
Unfortunately, due to the noise in RPA forces, as diagnosed by stable classical and path-integral simulations in the revPBE-D3 $NPT$ ensemble at 220\,K in Fig.~\ref{fig:rpa_simulations_ice_Ih}(d), we report unstable simulations in the $NPT$ ensemble at 220\,K and 1\,bar.
On the other hand, we can perform stable simulations in the $NVT$ ensemble at 220\,K and compare the predicted pair correlation function with the experiment in  Fig.~\ref{fig:rpa_simulations_ice_Ih}(c).
We report an overall good agreement with the experimental pair correlation function and with the revPBE-D3, modulo the over-structuring of the first and second peaks. 
Although quantum nuclear motion at 220\,K is expected to broaden the first and second peaks but not sufficiently enough to explain the extent of static disorder in experiments. 
The non-zero probability in the $3 - 4~\AA{}$ range could arise from defect migration at the grain boundaries in the power sample. 
Alternatively, the disagreement for short distances (or large reciprocal space vectors) could be an artefact of the empirical potential structure refinement~\cite{bowron_experimentally_2008} used to analyze experimental data.

\section{Conclusions}
\label{s:conclusions}

In summary, we explore the accuracy, extrapolation power, and data efficiency of the MACE architecture for predictive finite-temperature sublimation enthalpies of ice polymorphs.
In doing so, we present a simple workflow for first-principles quality studies of a polymorph of a molecular compound at a given temperature and pressure. 
First, we perform a short GGA level first principles $NPT$ MD simulation with coarse convergence parameters. 
Second, we randomly sample configurations and perform single-point total energy, force, and stress calculations with converged parameters to determine the appropriate choice of electronic structure theory.
Third, we fit MACE MLPs and perform simulations in the $NPT$ ensemble to calculate the density and the average potential energy. 
To estimate the sublimation enthalpy, we follow the same steps for the gas phase molecule. 
Finally, as an optional step, we perform $DFT$ calculations on the $NPT$ sampled configurations to estimate DFT-level thermodynamic quantities using statistical reweighting for direct testing. \\

Training a MACE model by finetuning the parameters of the pre-trained \MACEMP{} model, as opposed to training it from scratch, results in improved accuracy and data efficiency. 
Only 50 to 100 training structures sampled for a given $T, P$ condition are needed to achieve sub-kJ/mol and sub 1\,\%  agreement on the average energy and density, respectively, against the reference DFT $NPT$ ensemble.
Exploiting the accuracy and low data requirement of our approach, we develop an RPA-quality machine learning model for simulating ice Ih in the $NPT$ ensemble.
Our RPA simulations demonstrate an overall good agreement with the experimental density and pair-correlation functions and an improvement beyond DFT. 
At the same time, the noise in RPA training forces compromises the MLP's data efficiency and robustness compared to the DFT level.
Our work highlights the importance of tightly converged electronic structure theory training data, particularly at correlated levels.  \\

In future work, we aim to improve the accuracy of our approach and applicability to more complex systems.
These improvements include using MACE-MP-0 to sample the initial configurations to eliminate the computationally demanding first-principle MD for the initial dataset sampling, performing path integral simulations for initial sampling and for predicting quantities inclusive of quantum nuclear effects, and improving the data efficiency of the approach by pooling training configurations of various polymorphs.
With these developments, we foresee predictive sublimation enthalpy and physical property predictions for molecular crystals at the accuracy of correlated electronic structure and path integral molecular dynamics.

\section{Acknowledgements}

We thank Kit Joll, Keith Buttler, Sander Vandenhaute, and Michele Ceriotti for insightful discussions and their comments on the manuscript.
V.K. acknowledges support from the Ernest Oppenheimer Early Career Fellowship and the Sydney Harvey Junior Research Fellowship, Churchill College, University of Cambridge. 
J.L. thanks the Simons Foundation Postdoctoral Fellowship.
A.M. and X.R.A. acknowledge support from the European Union under the ``n-AQUA" European Research Council project (Grant no. 101071937).
I.B. was supported by the Harding Distinguished Postgraduate Scholarship.
We are grateful for computational support from the Swiss National Supercomputing Centre under projects s1209 and s1288, the UK national high-performance computing service, ARCHER2, for which access was obtained via the UKCP consortium and the EPSRC grant ref EP/P022561/1, and the Cambridge Service for Data Driven Discovery (CSD3).

\end{document}